\documentclass[aps,prl,preprint,superscriptaddress]{revtex4-1}
\usepackage{graphicx}
\begin{document}
\title{Effects of developmental variability on the dynamics and self-organization of cell populations.}

\author{Kaumudi H Prabhakara}
\affiliation{Laboratory of Atomic and Solid State Physics, Department of Physics, Cornell University, Ithaca, USA}
\affiliation{Max-Planck Institute for Dynamics and Self-Organization, G{\"o}ttingen, Germany}
\author{Azam Gholami}
\affiliation{Max-Planck Institute for Dynamics and Self-Organization, G{\"o}ttingen, Germany}
\author{Vladimir S Zykov}
\affiliation{Max-Planck Institute for Dynamics and Self-Organization, G{\"o}ttingen, Germany}
\author{Eberhard Bodenschatz}
\affiliation{Laboratory of Atomic and Solid State Physics, Department of Physics, Cornell University, Ithaca, USA}
\affiliation{Max-Planck Institute for Dynamics and Self-Organization, G{\"o}ttingen, Germany}
\date{\today}

\begin{abstract}
We report experimental and theoretical results on spatiotemporal pattern formation in cell populations, where the parameters vary in space and time due to mechanisms intrinsic to the system, namely \textit{Dictyostelium discoideum (D.d.)} in the starvation phase. We find that different patterns are formed when the populations are initialized at different developmental stages, or, when populations at different initial developmental stages are mixed. The experimentally observed patterns can be understood with a modified Kessler-Levine model that takes into account the initial spatial heterogeneity of the cell populations and a developmental path introduced by us, i.e., the time dependence of the various biochemical parameters. The dynamics of the parameter agree with known biochemical studies. Most importantly the modified model reproduces not only our results, but also the observations of an independent experiment published earlier. This shows that pattern formation can be used to understand and quantify the temporal evolution of the system parameters. 
\end{abstract}

\maketitle

\section{Introduction}
Pattern formation has been broadly studied in living and non-living systems over a large range of scales \cite{CrossHohenberg,Meron,CrossGreenside}, e.g.,  animal coats, shells, butterfly wings \cite{murray}, in dynamics of cardiac tissues \cite{heart}, in chemical reactions like the Belusov-Zhabotinsky reaction \cite{Winfree}. In some of these systems, for various cases, the dynamics change with time, leading to the formation of different patterns. In living systems the most common reason is the variation of gene expression levels. For example, during the embryogenesis of the fruit fly Drosophila, the activation of different genes at different stages causes the spatial patterning \cite{Meinhardt}. While these patterns can be visualized easily, correlating these patterns with the changing dynamics of the system is challenging.

One system that shows temporally varying patterns and self-organization is a starving population of \textit{Dictyostelium discoideum} (\textit{D.d.}). Ever since \textit{D.d.}'s behavior was described \cite{Raper,Bonner}, it has been extensively studied to explore its chemotaxis \cite{devreotes-chemotaxis,gerisch-chemotaxis}, pattern formation \cite{gerisch-waves,alcmonk}, self-organization \cite{Newell1977,Dormann}, multi-cellularity \cite{bonner1950}, development \cite{Klein1467} etc. Its natural environment is the soil, where it feeds on bacteria. The cells exhibit social behavior when they begin to starve \cite{kessinbook}. They secrete a chemical called cyclic adenosine monophosphate (cAMP) as a response to starvation. The secreted cAMP diffuses through the surrounding medium. The neighboring cells detect the cAMP through their membrane bound receptors and, with the help of the enzyme adenyl cyclase, secrete more cAMP in response. Some of the cAMP molecules are degraded by an enzyme called phosphodiesterase, which is present in intracellular, extracellular, and membrane bound forms \cite{parentpde}. The response of the cells to cAMP's passage through the external medium can be visualized with dark field optics, either as spirals or as targets \cite{alcmonk,Devreotes-imaging,Tomchik,Siegert}. After starving for about $5 \ h$, the cells respond to the external cAMP by migrating towards a higher concentration of cAMP - a process called chemotaxis. About $10^5$ cells aggregate to form mounds. These mounds then form multi-cellular slugs that scout for food. On failing to find nutrients, the slug develops into a fruiting body; the cells that form its stalk die and the cells at its top become spores \cite{Weijer-review}.

The developmental process from starvation to forming fruiting bodies takes about $24 \ h$. Throughout this time, the cells continually undergo changes. For example, at different starvation times, different receptors of cAMP are expressed \cite{Kim-car}, the cells secrete different amounts of cAMP \cite{Klein-aca,Klein-aca-2}, the expression of the amount of the degrading enzyme, phosphodiesterase, varies \cite{riedel-gerisch, malchow-pde}, an inhibitor of phosphodiesterase is expressed at later stages of the development \cite{Yeh}, etc. These changes are enabled by the expression of the respective genes \cite{pde-genexpression,aca-regulation}. It is well known that expression of many genes are controlled by the number of cAMP pulses they received \cite{KimmelFirtel,Chen-dev}. Therefore, as the cells starve, they undergo a continuous process of development. An important question is whether these developmental changes influence the patterns formed by starving populations of \textit{D.d.}

Various models have been proposed to explain the formation of patterns in starving populations of \textit{D.d.} \cite{MG-original,Tyson,Falcke,Goldbeter,PalssonCox,KL,LevinePNAS,Sawai}. The models describe mechanisms through which spirals could be generated in a system resembling populations of \textit{D.d.} One of earliest models was proposed by Martiel and Goldbeter \cite{MG-original}. This model considers some of the biochemical reactions occurring in the system. The dynamics are reduced to three variables, corresponding to intracellular cAMP, extracellular cAMP and the fraction of occupied receptors. This model was extended by Tyson et al. \cite{Tyson} to reproduce spatiotemporal patterns like spiral waves. It was further modified by Falcke and Levine \cite{Falcke} to include genetic variability of the membrane receptor of cAMP. The latter was achieved by introducing a new variable, which is the ratio of the total receptor concentration to the initial concentration. However, this modification assumes that the amounts of adenyl cyclase, intracellular phosphodiesterase and extracellular phosphodiesterase increase monotonically with this new variable. Further, to reflect the temporal changes that occur during the development of \textit{D.d.}, Lauzeral et al. \cite{Goldbeter} modified the model to incorporate time variation in adenyl cyclase and rate of degradation. However the time dependence of the parameters does not match previous biochemical experiments \cite{Klein-aca,Klein-aca-2,riedel-gerisch, malchow-pde}. A further modification was proposed by P\'{a}lsson and Cox \cite{PalssonCox}, in which the amount of phosphodiesterase in the system  was regulated in two ways: by having a random initial distribution of the amount of phosphodiesterase, and by gradually decreasing the amount of phosphodiesterase with time to account for the increasing levels of the phosphodiesterase inhibitor. Next, they tested this experimentally, by observing patterns formed by mutants lacking the inhibitor \cite{coxpdi}. They found that cells deficient in the gene of the inhibitor failed to form well developed spirals and formed smaller aggregates.

Another well-established model was proposed by Kessler and Levine \cite{KL}. This model mimics the behavior of \textit{D.d.} by placing ``bions" on a 2D grid. This is a cellular automata type of model, where each grid point, i.e., each bion, is governed by pre-determined rules. The bions are initially excitable. If the cAMP concentration surrounding them is higher than a threshold, the bions start to emit cAMP. Then they enter an absolute refractory phase where no secretion of cAMP is possible. After spending a specified amount of time in this phase, they revert back to being excitable. This model was also successful in reproducing spiral waves with suitable initial conditions. To make the model more realistic, Levine et al. \cite{LevinePNAS} modified it by introducing genetic feedback mediated by the pulses of cAMP. This was achieved by introducing a new variable called the excitability, which is coupled to the amount of cAMP, and in turn couples to the excitation threshold. This model was also able to simulate aggregation by including cell movement after the establishment of spirals. It was later slightly modified and used to account for patterns formed by populations deficient in certain genes of the signal transduction pathway \cite{Sawai}. Using a simpler form of cellular automata, another study numerically investigated the effects of variability \cite{hutt} by simulating a system that models the spread of epidemics. They focused on the parameter corresponding to the excitability of the cells and the parameter regulating the impact of ``infected" cells on the neighboring cells. To introduce variability, they assumed that these two parameters have a high value and a low background value. Each bion was assigned one of the two values for both parameters. The fraction of bions with the high value of the parameters was varied. This resulted in a variety of patterns. However, it is difficult to draw parallels with \textit{D.d.} because the model used is far from the biochemical processes involved in \textit{D.d.} Further, temporal variations of the parameters are not considered.

Although each of these models and their modifications explain various experiments, many results on pattern formation and selection in \textit{D.d.} need further explanation. For instance, a study \cite{Dormann-cAR} showed that mutants expressing different kinds of cAMP receptors produce different kinds of patterns. In another experiment Lee et al. \cite{resetting} perturbed pattern formation by spraying a mist of cAMP on populations showing spirals. Depending on how long after the starvation the mist was sprayed, the populations either managed or failed to re-produce spirals. None of the models described above reproduced all the experimental results presented below.

In this work, we show that cell development and heterogeneity play an essential role in pattern formation of \textit{D.d.} As stated before, while the populations starve, their gene expression levels vary. Therefore, in order to introduce genetic heterogeneity and variability, we starve populations for different initial durations (see Methods). In one set of experiments, we observe the patterns formed by single populations that have been starved for different initial durations. In the second set of experiments, we mix two populations containing equal number of cells that have been starved for different initial durations. We analyze the patterns formed in these experiments and simulate them with the modification of the Kessler-Levine model used by Sawai et al. \cite{Sawai}. We show that spatial heterogeneity of cell density and parameters combined with a specific temporal evolution of the system parameters (developmental path) semi-quantitatively reproduce the experimental results. Moreover, our modified model also reproduces the results of cAMP resetting experiments carried out by Lee et al. \cite{resetting}.

\section{Results}
\subsection{Single populations}
\begin{figure}
\setlength\belowcaptionskip{-1ex}
 \includegraphics[scale=0.5]{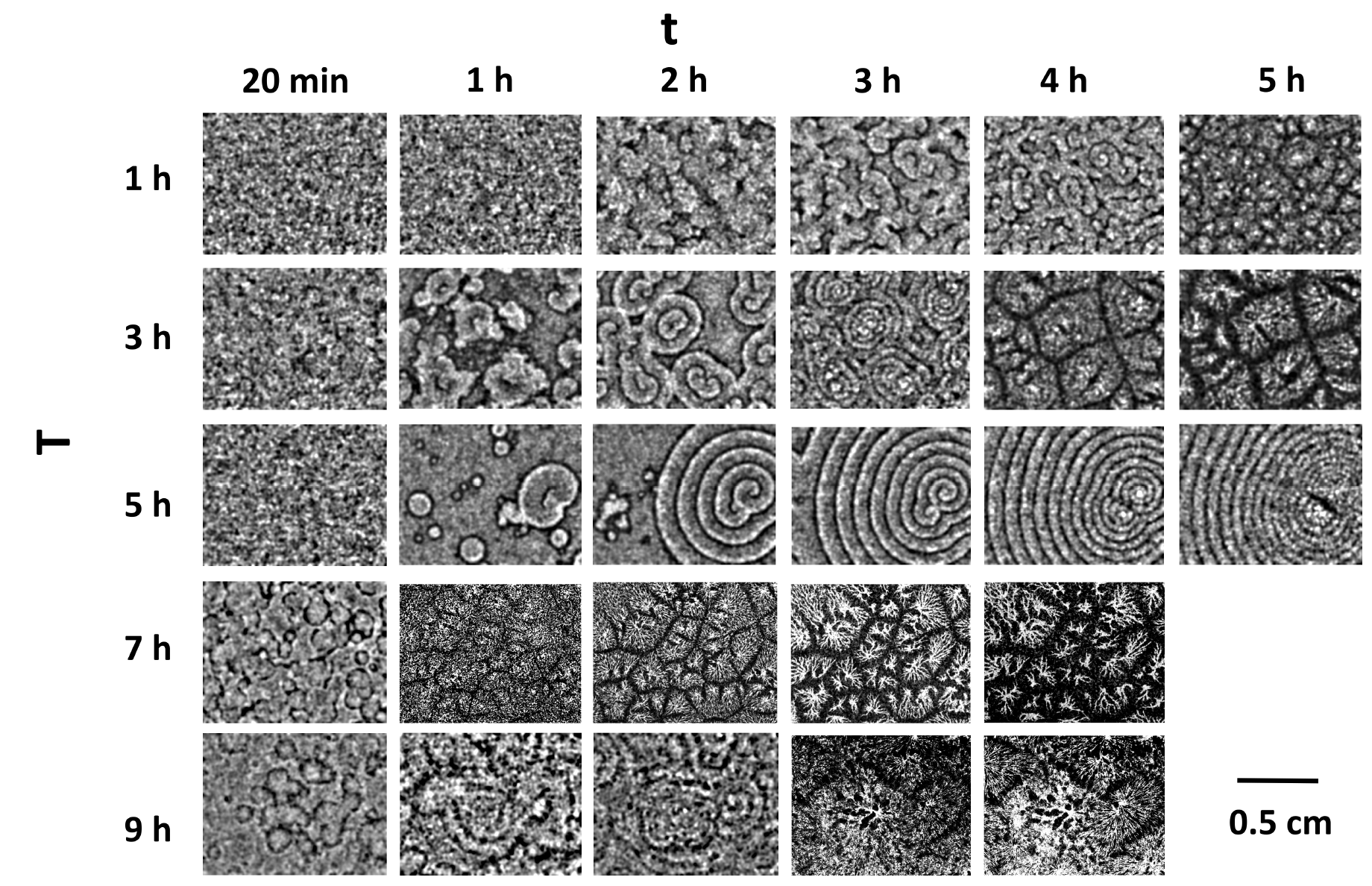}%
 \caption{Sample patterns obtained in experiments for different initial starvation times T as a function of the observation time t. \label{singpopexp}}
 \end{figure}

To systematically study the effect of development on pattern formation, we starved populations of AX2 wild type cells for different durations. After this initial starvation, the populations were plated on a Petri dish and observed under a dark field set-up. (See Methods.) This ensured that the populations started with different expression levels of the biochemical parameters. We observed different patterns depending on the initial starvation times as shown in figure \ref{singpopexp}. When populations were starved for short durations (less than $4 \ h$), they formed numerous small spirals. These spirals started from broken wave segments. Populations starved for intermediate times (between $4 \ h$ and $6 \ h$) formed fewer spirals. The patterns of these populations started differently - targets appeared at first, which were then replaced by spirals (compare supplementary movies S1 (https://youtu.be/oYRF7BaaaJY) and S2 (https://youtu.be/kT0R3wNbiro)). Populations starved for longer times (more than $7 \ h$) only formed targets and began streaming soon after the formation of the initial patterns. 
Further important observations should be noted. Consider, for example, the patterns formed by a population starved initially for $3 \ h$. Figure \ref{singpopexp} shows the patterns observed after another $3 \ h$ of signaling. At this point, the population had starved for a total of $6 \ h$. Now consider instead, a population initially starved for $5 \ h$. Figure \ref{singpopexp} shows the patterns observed after $1 \ h$ of signaling. At this point in time both populations had been starved for a total of $6 \ h$. Please note that the patterns of the two populations strongly differ, clearly showing the importance of the developmental path and the associated gene expression on the pattern formation of \textit{D.d.}
\begin{figure}
\setlength\belowcaptionskip{-1ex}
 \includegraphics[scale=0.2,angle=270]{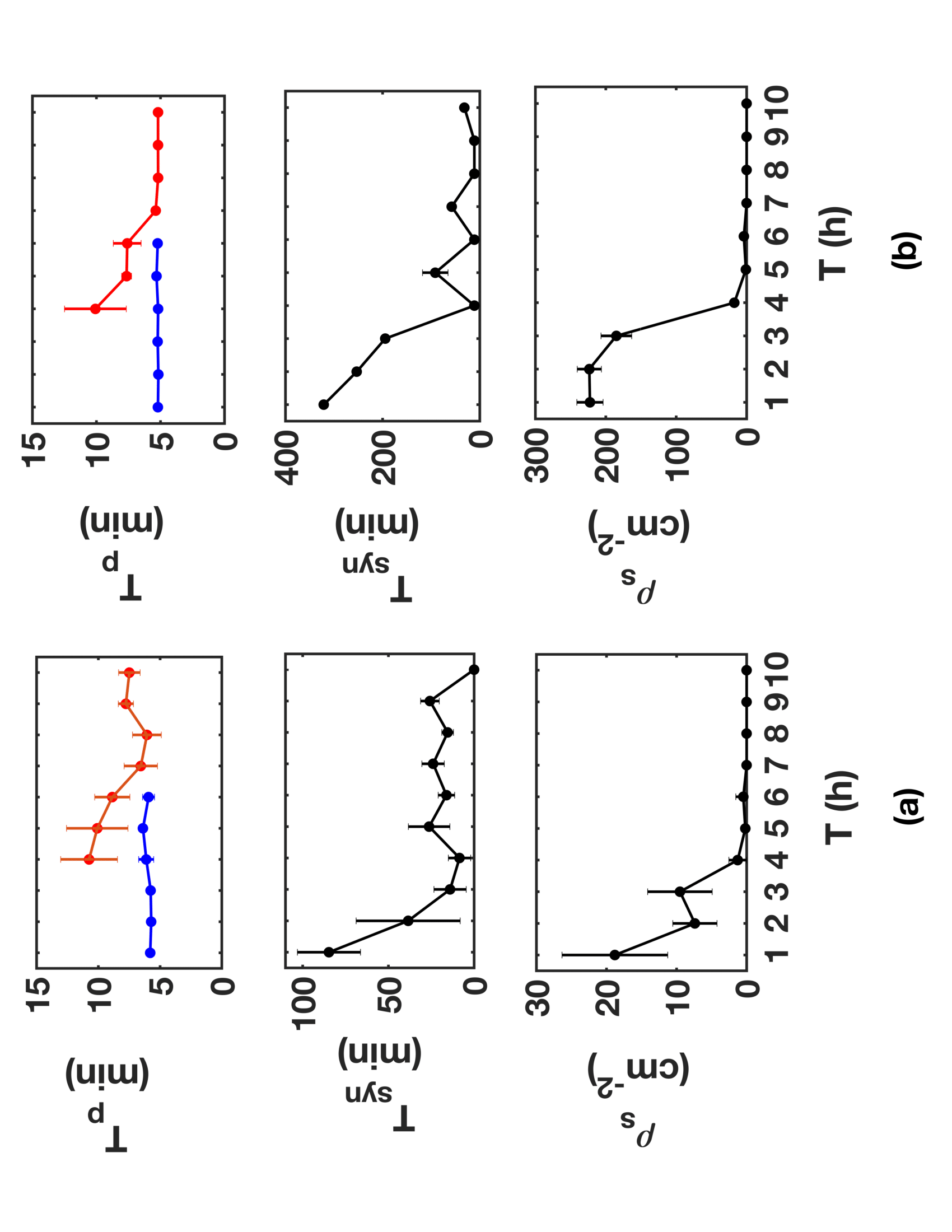}%
 \caption{Parameters evaluated from the patterns (a) from experiments and (b) from simulations. The red lines (blue lines) in the top panels denote the oscillation period, $T_p$ of target (spirals). The central panels show the time taken for onset of synchronization, $T_{syn}$, and the bottom panels show the spiral density, $\rho_s$. The error bars correspond to the standard deviation of the experiments and simulations. Different runs of the simulations (6 for each starvation time) were achieved by having different random initial conditions. See also Methods. \label{exp_sim}}
\end{figure}
To quantify the patterns, we use three order-parameters from the experimental data: oscillation period $T_p$, onset of synchronization $T_{syn}$, and spiral density $\rho_s$. The Methods Section describes the exact procedure for the data analysis. Figure \ref{exp_sim} shows the variation of these quantities as a function of initial starvation time T of the populations. Spirals have an oscillation period between 5 $min$ and 6 $min$, irrespective of initial starvation time. The period of the spirals decreases with observation time and becomes approximately constant. Between $4 \ h$ and $6 \ h$ of initial starvation time, we observed a co-existence of spirals and targets. In this range, targets always had a longer oscillation period compared to spirals. This shorter period of the spirals enables them to take over the patterns.  Spirals had a very small variation in oscillation period compared to targets as shown by the error bars. This indicates that spirals always have a fixed frequency of oscillation, whereas the frequency of target patterns varies. This is expected and well known from theoretical studies \cite{Keener-tyson}. 

Next, let us consider the time it takes for the populations to exhibit a synchronization of cell dynamics. Such a time is quantified by the onset of synchronization plotted in the figure \ref{exp_sim}. With increasing initial starvation time, populations form patterns faster. The synchronization time decreases as initial starvation time increases. 

Finally, we characterize the patterns by the number of spirals formed in a given area. It can already be seen in figure  \ref{singpopexp} that the number of spirals decreases with increasing initial starvation time. Figure \ref{exp_sim} quantifies this: the spiral density reaches a minimum at $5 \ h$ of initial starvation time, and then increases slightly at $6 \ h$ (compare supplementary movies S3 (https://youtu.be/w-EDe9NIeN0) and S4 (https://youtu.be/LGHIlz1oU24)). At $5 \ h$ of initial starvation time, due to the minimum in spiral density, very large spirals are formed, as is seen in figure \ref{singpopexp}. (Supplementary movie S3 (https://youtu.be/w-EDe9NIeN0).)
\begin{figure}
\setlength\belowcaptionskip{-1ex}
 \includegraphics[scale=0.5]{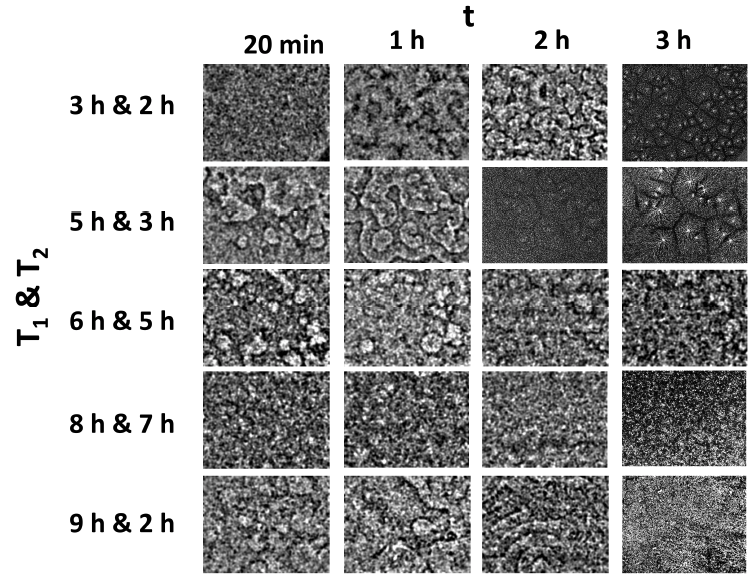}%
 \caption{Sample patterns obtained in experiments for different mixtures as a function of the observation time t. $T_1$ and $T_2$ denote the initial starvation times of the two populations. The width of each panel is $1.8 \ cm$.\label{mixexp}}
 \end{figure}
\begin{figure}
\setlength\belowcaptionskip{-1ex}
 \includegraphics[scale=0.5]{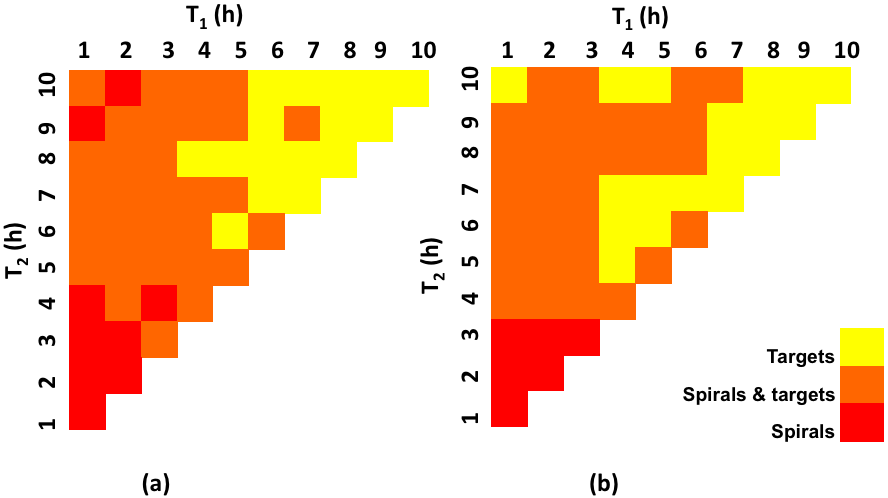}%
 \caption{Phase diagram containing the classification of the patterns obtained in (a) experiments and (b) simulations. $T_1$ and $T_2$ denote the initial starvation times of the two populations. \label{phasemap}}
 \end{figure}
\subsection{Binary population mixtures of different developmental stage}
So far we have seen that developmental time plays an important role in pattern formation of \textit{D.d.} Next, we study the effect of developmental heterogeneity on pattern formation. We starved populations containing equal number of cells for different initial starvation times (as we described under Methods) and then mixed them before observation. This way, cells with different gene expression levels interacted with  each other.  In the experiments we used all possible binary mixtures of populations initially starved between $1 \ h$ and $10 \ h$ (in $1 \ h$ intervals). For convenience, we call the populations initially starved for shorter (longer) durations, the younger (older) populations. The mixtures formed a variety of patterns as shown in figure \ref{mixexp}.

A phase diagram (figure \ref{phasemap}), summarizes the patterns observed. The two axes correspond to the initial starvation times of the two populations. The hypotenuse of the phase diagram corresponds to single populations. In the lowest part of the phase diagram, where the populations are initially starved for less than $4 \ h$ each, numerous small spirals were observed. These spirals formed from broken wave segments. In the top left part of the phase diagram, where population 1 was initially starved for less than $5 \ h$ and population 2 was initially starved for more than $5 \ h$, spirals co-existed with targets. In these mixtures, the populations formed targets first and then spirals. But the spirals, having a lower period took over the targets. Older populations (initially starved for longer than $8 \ h$) that did not form spirals on their own, formed spirals when mixed with a younger populations. The patterns appeared faster in the mixtures than they would have in the younger populations only. These observations indicate that the older populations are important contributors to the pattern formation process. In the top right part of the phase diagram, both the populations were initially starved for longer than $5 \ h$. These mixtures mostly failed to form spirals. After a few oscillations of the targets, they immediately began to stream. As in the case of single populations, the images of these mixtures had a low signal to noise ratio, which made measurement of spiral density less reliable.

\subsection{Simulations}
To understand our experimental results, we rely on the model originally proposed by Kessler and Levine \cite{KL} and modified by Sawai et al. \cite{Sawai}. Let us first summarize this model and its parameters before we show that our experimental results necessitate further modifications. The model simulates the collective dynamics of bions placed on a 2D grid. The dynamics of the bions obeys the following equations: 

\begin{equation} \label{modeleq1}
\dot{C}_{i,j} = D \cdot L(C_{i,j})  -\gamma \cdot C_{i,j} + \theta \cdot C_{rel}
\end{equation}
\begin{equation} \label{modeleq2}
\dot{E}_{i,j}=\eta +\beta \cdot C_{i,j} ; \quad E_{i,j} \leq E_{max}
\end{equation}
\begin{equation} \label{modeleq3}
C_{i,j}^{thresh}= \Bigg[C_{max} - A\frac{\tau}{\tau+T_{ARP}}\Bigg](1-E_{i,j}), 0 < \tau < T_{RRP}
\end{equation}
\\
In equation (\ref{modeleq1}), L is the discrete Laplacian operator and $\theta$ is a step function - it is 1, if $C_{i,j}$ $\geq$ $C_{i,j}^{thresh}$ and 0 otherwise. (It is always zero in the absolute refractory phase, as described below.) Initially, the bions are excitable. If the concentration $C$ of cAMP exceeds a threshold $C^{thresh}$, the bions emit cAMP at a rate $C_{rel}= 300/min$ for $1 \ min$. Immediately after emission, they enter an absolute refractory state, where they are incapable of emitting cAMP for a fixed time, $T_{ARP}= 2 \ min$. Next, the bions go into a relative refractory state, which has a high threshold for emission of cAMP. This state also lasts for a fixed time $T_{RRP}= 7 \ min$. The threshold required for cAMP release decreases throughout this state (as $\tau$, the time spent in the relative refractory phase, increases from 0 to $T_{RRP}$) and reaches a minimum at the end of this state (in accordance with equation (\ref{modeleq3})), after which the bions are excitable again and the cycle continues. The cAMP is degraded by an enzyme at a rate $\gamma = 8/min$. It is important to note that the degradation rate is the cumulative effect of phosphodiesterase and its inhibitor. The excitability of the cells $E$, increases autonomously, with $\eta = 0.0001/min$ and is proportional to $C$ with a coupling $\beta = 0.005/min$. It saturates at a fixed value $E_{max}=0.93$ (equation (\ref{modeleq2})). The time spent in the relative refractory phase ($\tau$) and the excitability ($E$), determine the concentration of cAMP required to cause emission of cAMP by a bion. Highly excitable bions (high $E$) have a very low threshold for initiating cAMP emission and can be easily excited to release cAMP, whereas bions with low excitability have a higher threshold and are not easily excitable. $C_{max}=100$ is a constant describing the maximum value of the threshold required for emission of cAMP, D = 0.00138 $mm^2/min$ is the diffusion coefficient and the constant $A=(T_{RRP}+T_{ARP})(C_{max}-C_{min})/T_{RRP}$ is chosen so that the threshold is maximum when the bions enter the relative refractory state and minimum when they leave it. 

All bions fire randomly with a probability of 
$10^{-4}$
. The simulations were initialized with $E_0=0.5$ and $C=0$ everywhere.  A small fraction of bions were set to emit cAMP. Although the fraction used in Sawai at al. \cite{Sawai} is unknown, we found that $5$ emitting bions randomly chosen out of 333 x 333 bions placed on a square grid reproduced their results. A bion that has already emitted cAMP at-least once, and has not emitted cAMP for about 15 $min$ (the quiescence time) is forced to emit cAMP at the end of the quiescence time. The rules of the fixed durations of the refractory phases and the saturation of the excitability are implemented at each bion separately from these equations. The threshold required for cAMP release, $C^{thresh}$ is calculated at each bion at each time step. The equations \ref{modeleq1}-\ref{modeleq3} are solved using the explicit Euler method with a grid size of $0.06 mm$ and time step of $0.01 min$.
\begin{figure}
\setlength\belowcaptionskip{-1ex}
 \includegraphics[scale=0.15]{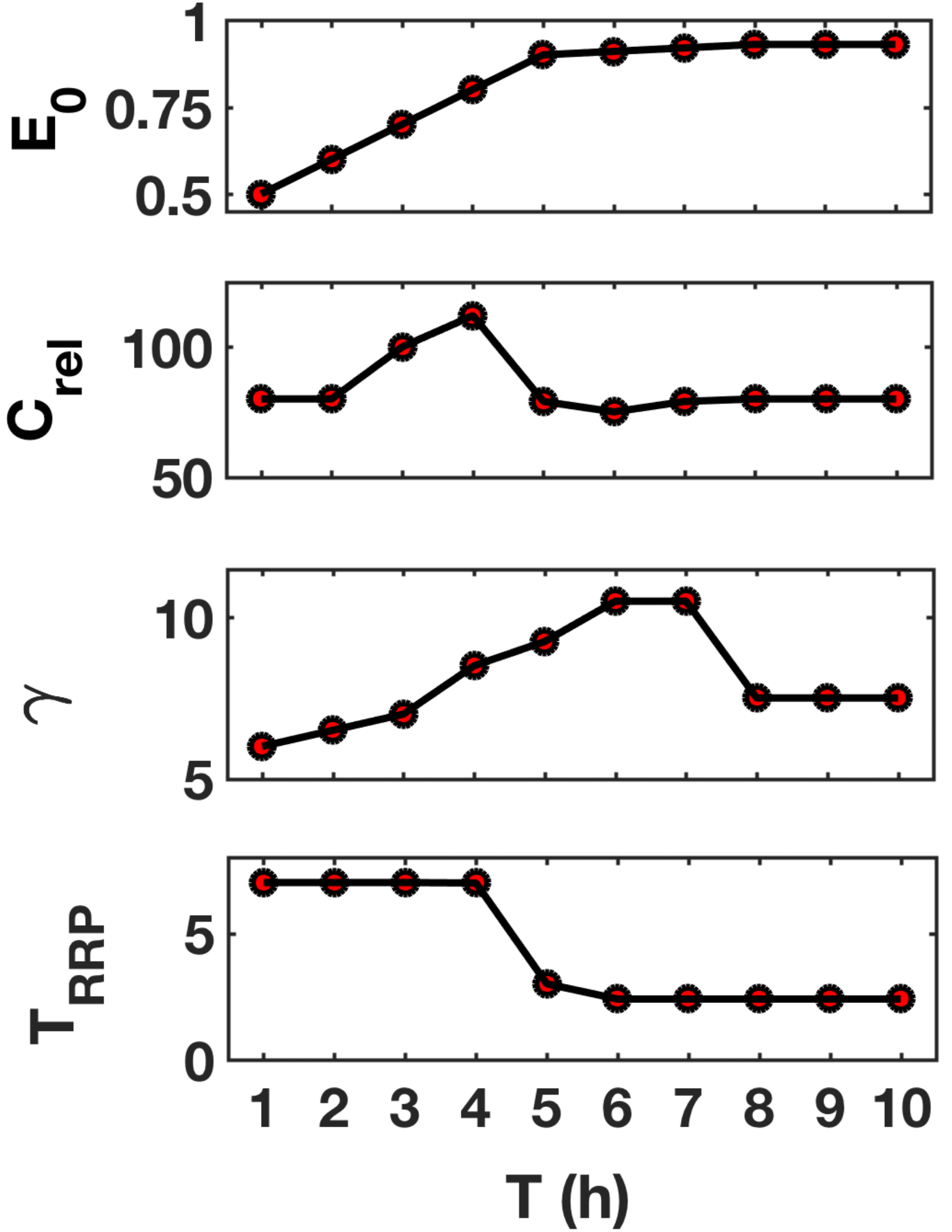}%
 \caption{The temporal variation of the parameters used in the modified model. $E_0$ is the initial excitability for equation (\ref{modeleq2}). $C_{rel}$ denotes the amount of cAMP released by the cells. $\gamma$ denotes the rate of degradation. It has units of $min^{-1}$. $T_{RRP}$ denotes the duration of the relative refractory period. Its unit is $min$. For $E_0$, only the initial value is taken from the top panel. With time, the variable $E$ evolves according to equation (\ref{modeleq2}). However, the other three parameters start and evolve with time according the respective curves shown here. \label{modelparams}}
 \end{figure}

In our simulations of the above model, we used populations at different initial starvation times  \cite{Sawai}, i.e., different initial starvation times correspond to different initial excitabilities of the bions. We studied both homogeneous populations and binary mixtures.  In the latter case, each bion consisted of a 1:1 mixture of populations with different initial starvation times.  The excitabilities, the cycle through the different phases, and the calculation of the threshold for cAMP release was conducted independently for each initially starved population. To our surprise, the model \cite{Sawai} did not reproduce the experimentally observed co-existence of spirals and targets for pure and mixed populations. Additionally, for large initial starvation times, the oscillation period of targets found in the simulations was about twice as long as those observed in experiments. Although the model was able to successfully capture the behavior of previous experiments  by Sawai et al. \cite{Sawai}, it is clearly insufficient to capture our experimental results. Thus, the model has to be modified. In our work, we conserve the structure of the model and introduce time dependence in some of its parameters.


Our first conjecture was that the oscillations period could depend on the time spent in the refractory period. Studies have shown that in \textit{D.d.}, the refractory period for the relay of signals decreases from about 7 $min$ to about 2 $min$ during development \cite{Durston-refract}. We capture this by decreasing $T_{RRP}$ with starvation time (figure \ref{modelparams}). With this modification the oscillation period of targets agreed with the experimentally observed value of around 6 $min$. However, this modification alone did not reproduce the co-existence of spirals and targets and further modifications were necessary. Previous studies have shown that the activity of phosphodiesterase increases with starvation time in suspension of  \textit{D.d.} \cite{riedel-gerisch, malchow-pde}. To capture this, we modified the model to include a temporal variation of the rate of degradation of cAMP ($\gamma$ in equation (\ref{modeleq1})) as shown in figure \ref{modelparams}. Please note that this choice agrees with these biochemical studies \cite{riedel-gerisch, malchow-pde}. Furthermore, it is known that cells release different amounts of cAMP as they starve  \cite{Klein-aca,Klein-aca-2}. Therefore, we varied the amount of cAMP emitted by the cells ($C_{rel}$ in equation (\ref{modeleq1})) with starvation time in a manner compatible with these previous studies (figure \ref{modelparams}). By varying these parameters, we are effectively varying the expression levels of the genes that regulate these parameters.
\begin{figure}
\setlength\belowcaptionskip{-1ex}
 \includegraphics[scale=0.5]{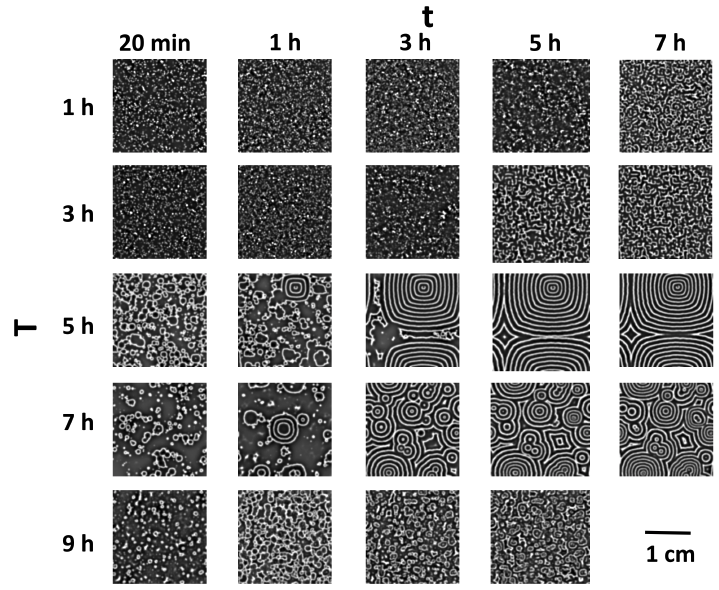}%
 \caption{Sample patterns obtained in simulations for different initial starvation times, T as a function of the observation time t. \label{singpopsim}}
 \end{figure}
In the simulations of the above model, the initial excitability was chosen according to figure \ref{modelparams}. Furthermore, to capture the initial conditions in the experiment as closely as possible, we included the spatial heterogeneity in the cell distribution in our model. See Methods for the details. 
\begin{figure}
\setlength\belowcaptionskip{-1ex}
 \includegraphics[scale=0.6]{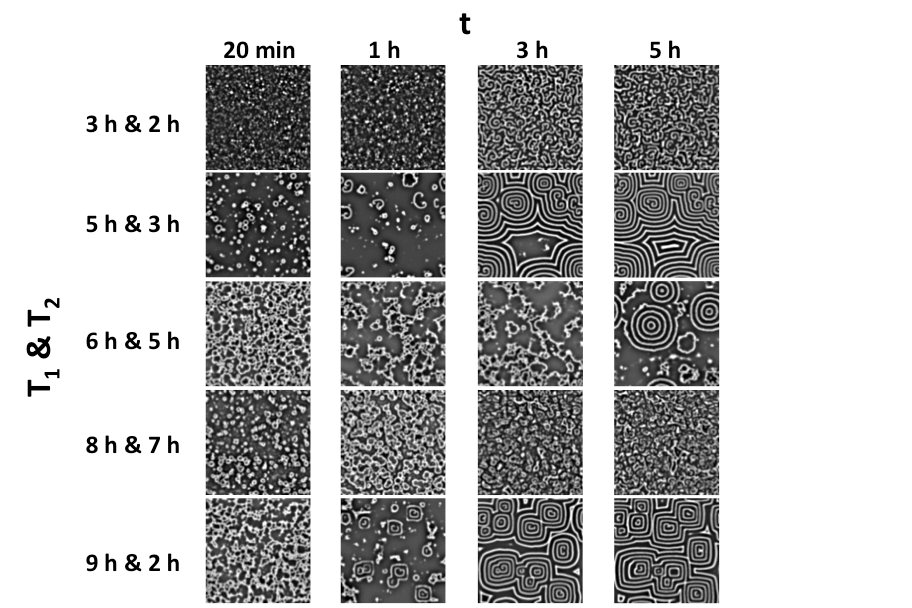}%
 \caption{Sample patterns obtained in simulations of different mixtures as a function of the observation time t. $T_1$ and $T_2$ denote the initial starvation times of the two populations. The length of each panel is $2 \ cm$. \label{mixsim}}
 \end{figure}

With these changes, i.e., the modification of the initial conditions and the time variation of the three parameters, the model semi-quantitatively reproduces our experimental results. Representative patterns obtained in the simulations are shown in figure \ref{singpopsim} for single populations and in figure \ref{mixsim} for mixtures of populations. Simulation results for all the mixtures and single populations are summarized in the phase diagram in figure \ref{phasemap}b, which agrees very well with the phase diagram from the experiments (figure \ref{phasemap}a). To further quantify our simulation results, we measured the three order-parameters using the same methods used in the experiments (see Methods). As shown in figure \ref{exp_sim}b, the measured order-parameters agree well with those in the experiments (figure \ref{exp_sim}a) for all cases investigated. 
\begin{figure}
\setlength\belowcaptionskip{-1ex}
 \includegraphics[scale=0.5]{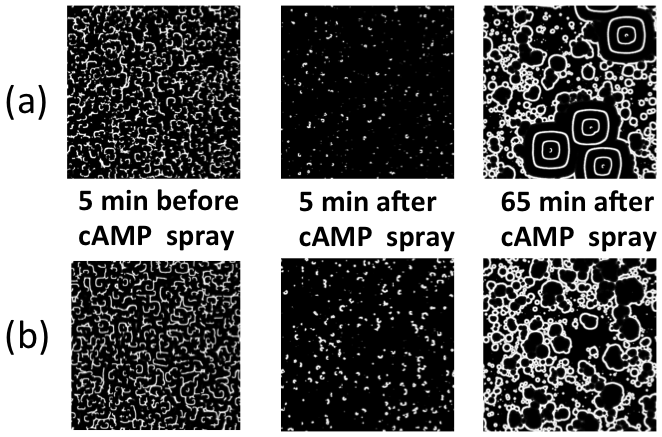}%
 \caption{Patterns obtained by simulating experiments by Lee et al \cite{resetting}. (a) Spirals re-emerge within an hour when the cAMP is sprayed at $4 \ h$ after start of simulations. (b) Only targets appear when cAMP is sprayed after $8.3 \ h$ of simulation. The length of each image is $2 \ cm$. \label{leeexp}}
 \end{figure}

At this point, one could argue that the model was modified to explain our experimental results and may not have any predictive value. Therefore it is imperative to test this model in an independent setting. In their seminal paper, Lee et al. \cite{resetting} conducted experiments on the pattern formation of \textit{D.d.} populations starving on agar. They observed targets and spirals if the system developed without any perturbation. However, when the cells were reset by spraying a mist of cAMP solution on the patterns, they showed different behavior depending on when the mist was sprayed. If cAMP was sprayed before $6 \ h$ of starvation, spirals and targets emerged; but if it was sprayed after $6 \ h$, only targets emerged. These experimental findings have been simulated by Falcke and Levine \cite{Falcke} using a generalized version of the Martiel - Goldbeter model \cite{MG-original} which includes the coupling of the cAMP signaling to the expression levels of aggregation genes, and qualitatively explained in other works \cite{resetting,coxpdi}. Here, we test if the modifications we made to the Sawai et al. \cite{Sawai} model to explain our experimental results can also reproduce the experimental results of Lee et al. \cite{resetting}


To mimic the cAMP spray, for one time step we set the concentration of cAMP at all bions to a value of 20, which is about twice the maximum cAMP concentration of normal signaling. In all simulations, the populations had an initial starvation time of $3 \ h$. If we ``spray" cAMP $1.7 \ h$ after start of the simulation ($4.7 \ h$ starvation) we observe patterns consisting of spirals and targets, just as in the experiment. As we increased the time at which we sprayed cAMP, the number of spirals that re-formed decreased. When cAMP was sprayed later than $4 \ h$ of simulation ($7 \ h$ of starvation), no spirals were formed again. The results of the simulation for two cases are shown in the figure \ref{leeexp}. It should be noted here that the experiments by Lee et al. \cite{resetting} were conducted on 2\% agar, whereas we modified the model to match experiments conducted on a plastic Petri dish. We know \cite{agarpaper} that pattern formation and time taken for synchronization depend on the substrate. So, some differences between the patterns seen by Lee et al. \cite{resetting} and patterns reproduced by our modified model are to be expected. 

\section{Discussion}
Let us now discuss some possible implications of our results. We observe a change in the way spirals are formed at around $4 \ h$ of initial starvation. For initial starvation times lower than $4 \ h$, the spirals form out of pre-existing singularities or broken wave segments. While for initial starvation times between $4 \ h$ and $6 \ h$ spirals co-exist with targets, for initial starvation times longer than $6 \ h$, only targets are formed. A possible explanation of the transition in the way the spirals form can be found in our modified model. For equation (\ref{modeleq2}), we assumed that the initial excitability increases linearly with time with a saturation after $6 \ h$ of initial starvation (figure \ref{modelparams}). This implies that in the experiment where only targets are observed, the cells are highly excitable. These highly excitable cells should have a short refractory period (figure \ref{modelparams}) and a low threshold for cAMP emission (equation (\ref{modeleq3})). The low threshold, in turn, causes the cells to produce pulses of cAMP, if they detect even a small amount of cAMP in their surroundings. Furthermore, due to their short refractory periods, the cells can quickly become excitable again. Thus, the cells emit cAMP frequently, and there are few refractory cells to cause the wave break required to induce spirals, resulting in the formation of targets and a few large spirals. As the probability of encountering refractory cells decreases with increasing initial starvation times, only targets are formed.

It is important to note that when spirals formed by cells with lower excitabilities, they were sustained even when the cells reached high excitabilities. For example, a population with an initial starvation time of $1 \ h$ was able to maintain the spirals even after $5 \ h$ of starvation, when the cells become highly excitable, as seen in experiments and in simulations. The initial excitability of cells is therefore a very important factor in determining the kind of patterns formed. 


However, it is not just the refractory period that determines the spiral density. The combination of $\gamma$ and $C_{rel}$ is critical to obtain the right trend in spiral densities. For example, in experiments, we observed that the spiral density increased slightly after reaching a minimum at $5 \ h$ of initial starvation. Such trends can be achieved in simulations only by using a particular set of $\gamma$ and $C_{rel}$ values (see supplementary movies S5 (https://youtu.be/Mt1AfTDreSk) and S6 (https://youtu.be/8CQzvBzOHOc)). Therefore, the time variation of the three parameters, the refractory period, rate of degradation and amount cAMP released, is crucial in determining the kind of patterns observed. The degradation rate and the amount of cAMP released are controlled by the enzymes phosphodiesterase, phosphodiesterase inhibitor and adenyl cyclase. Therefore, it is the variation in the expression levels of the genes controlling these enzymes that is responsible for the pattern selection.

An interesting question about pattern formation in \textit{D.d.} is whether the dynamics are excitable or oscillatory. The Kessler-Levine model always considers the dynamics of the cells to be excitable. This is true for the populations starved for short periods of time. However, as discussed above, after about $6 \ h$ of initial starvation, the excitability of the cells reaches a maximum value. So the right hand side of equation (\ref{modeleq2}) is zero. Also, as mentioned earlier, the high excitability reduces the threshold of surrounding cAMP required to cause emission of cAMP (equation (\ref{modeleq3})). As a consequence cells immediately fire when they leave the refractory period ($T_{ARP}+T_{RRP}$), i.e., the cells are oscillatory.


Finally, heterogeneity is also an important factor for pattern formation. To reproduce patterns formed by older populations in simulations, it was essential to incorporate the initial spatial heterogeneity in the cell distribution --- cells in older population tend to form clusters, leaving empty spaces in between. Further, the importance of heterogeneity in parameters is best seen in mixtures. When older populations that formed targets on their own were mixed with younger populations, spirals were formed. The time taken to form these spirals is less than the time taken by the younger populations. Therefore, the synchronization of the population and the pattern selection depend on the heterogeneity of the parameters. 

Before concluding, a note about our simulations is pertinent. We tried various combinations of parameters, with and without time variation. Since these changes were manual, it was impossible to test all possible parameter combinations. We chose the three parameters that are known to change with starvation time and changed them in a manner compatible with previous biochemical experimental results and in a manner that best reproduced our experimental patterns for both single populations and mixtures of two populations. We do not claim this to be the only way or the best way to modify this model, however our choice of parameter evolution seems to work remarkably well. A mathematically rigorous method of parameter estimation is highly desired, but currently unavailable. 

Although the modified model semi-quantitatively reproduces all experimental results, it is far from perfect. The values of the onset of synchronization and spiral density don't match the experiments as shown in figure \ref{exp_sim}. Also, for the mixed populations, though the phase diagrams look similar, for short initial starvation times actual parameters like the spiral density are different from the experiments (quite similar to that of single populations). Another point to note is that in some cases, e.g. initial starvation time of $9 \ h$, the simulations produce small spirals after few hours of simulation time, while this is not the case for the experiments, where the cells have already streamed at that time. Thus, to classify the patterns in figure \ref{phasemap}b, we considered only simulations over a time interval corresponding to the experimental one. One possible way to fix this issue would be the introduction of cell motility in our modified model \cite{KL}.
\section{Conclusions}

We have studied the effects of development on pattern formation in \textit{D.d.} experimentally, and explained the results with a modified model based on \cite{KL,Sawai}. From our analysis of the observed patterns, we conclude that the selection of patterns formed by \textit{D.d.} depends on the temporal variability of the parameters, i.e., the initial excitability and the spatial heterogeneity of parameters and cell density. 

Recapitulating, pattern formation in \textit{D.d.} depends on how long the cells have been starved for. Older populations formed target patterns while younger ones formed spirals. The transition occured around $4 \ h$ of initial starvation. However, when old populations were mixed with younger populations, they formed spirals. We have quantified the patterns in terms of spiral density, synchronization time, and oscillation period. We have offered explanations for the variation of these quantities and the observed transition in the patterns. We found that our modified model describes the experimental results by Lee et al. \cite{resetting}. We have shown that developmental variability is vital for pattern selection since the variation in parameters is brought about by the developmental path. 


We can predict the time evolution of the properties of \textit{D.d.} since our modifications lead to a semi-quantitative reproduction of the experimental observations. First, the excitability increases with starvation and saturates to a maximum value. At this maximum value the dynamics of the system can be thought of as oscillatory, whereas the system is excitable for short starvation times. Second, the refractory period of the cells decreases with starvation time. Third, the amount of cAMP released by the cells and the degradation rate of cAMP have a peak in their values during starvation. We can also propose certain rules for pattern selection: highly excitable cells cannot form spirals, but form targets; but highly excitable cells can sustain spirals, if their initial condition was a spiral. Not only do these results provide insights into the temporal behavior of \textit{D.d.}, our results are a proof of concept that it is possible to deduce many properties of the system by analyzing the patterns only. Furthermore, since the mechanism of pattern formation in \textit{D.d.} is similar to the mechanism of pattern formation in heart tissues and various chemical reactions, our results imply that an analysis of patterns in other excitable systems could also provide important information about the system parameters. 

\section{Methods}
\subsection{Experiments}
AX2 cells were grown in HL5 medium (35.5g of Formedium powder from Formedium Ltd., England, per liter of double distilled water, autoclaved and filtered) at $22^0C$ and harvested when they became confluent. The cells were washed in phosphate buffer (2g of $KH_2PO_4$, 0.36g of $Na_2HPO_4.2H_2O$ per liter at pH 6, autoclaved) and centrifuged two times. The cells were then counted using a hemocytometer. Then they were poured into a conical flask and placed on a shaker for the desired initial starvation time. After the desired starvation time, they were centrifuged and diluted to a density of 4 x $10^6$ cells/ml with fresh phosphate buffer. For mixtures, each population contributed 2 x $10^6$ cells/ml, and were starved in separate conical flasks. They were then plated on a $8.6 \ cm$ diameter plastic Petri dish with 10 ml phosphate buffer. 

Dark-field optics was used to image the cell populations every 20 s  \cite{alcmonk,Devreotes-imaging,Tomchik,Siegert}. It consists of an LED ring lamp placed above the Petri dish and a CCD camera (QIClick-F-M-12 from QImaging) placed below the Petri dish. Parafilm was wrapped around the sides to prevent evaporation of the buffer. A table fan blew air above the dish to equilibrate the temperature. The entire room was dark during the experiment and maintained at $22^0C$. The single populations experiments were repeated three times. The imaged area was split into four parts and each part was analyzed separately. Each mixture experiment was performed either once or twice.
\begin{figure}
\setlength\belowcaptionskip{-1ex}
 \includegraphics[scale=0.65,]{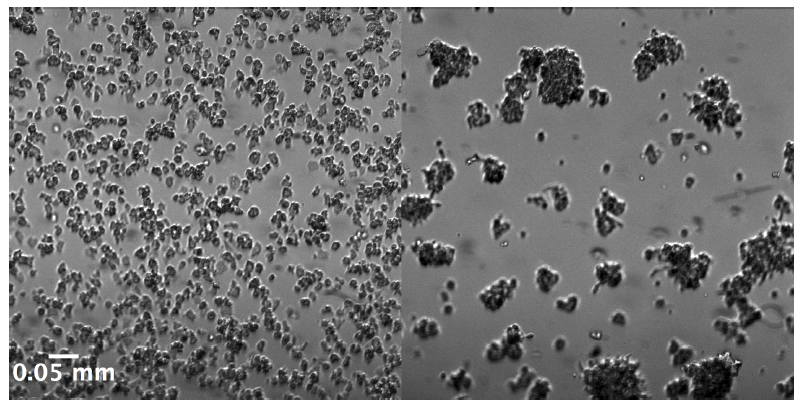}%
 \caption{The initial distribution of the cells observed experimentally. The left panel shows the cell distribution for an initial starvation time of $2 \ h$. The right panel shows the cell distribution for mixture of $10 \ h$ and $8 \ h$ initially starved populations.\label{celldist}}
\end{figure}
To find the spatial heterogeneity in the cell density, we experimentally analyzed the spatial distribution of cells for the different initial starvation times. We plated the starved cells at the same density on a Petri dish. After waiting for about 10 $min$ to allow the cells to attach to the surface, we imaged the dish at about 15 distinct regions using an inverted bright-field microscope, using a 10X objective. Figure \ref{celldist} exemplarily shows the initial cell distributions for two different initial starvation times. To determine the ratio of the area occupied by the cells to the area of the grid, i.e., the occupation ratio, we divided the images of about 600 x 600 $\mu m^2 $ into grids of size 60 x 60 $\mu m^2$ (corresponding to the size of the grids in the simulations). For these 600 x 600 $\mu m^2$ regions, we thus measured the spatial distribution of occupation ratios at 10 x 10 grid points (bions in the simulations). For each starvation time, we have about 15 such distributions. To cover the simulation region of 333 x 333 bions, we distribute these 10 x 10 distributions randomly on the simulation grid. We multiply the degradation rate $\gamma$ and cAMP released $C_{rel}$ for each bion with its respective occupation ratio, as the degradation rate and the amount of cAMP released  depend on the number of cells in each bion. For mixtures, we assume that each population makes up half the occupation ratio. This ensures that the effect of having clusters and sparsely occupied regions is incorporated into the simulation. 
\begin{figure}
\setlength\belowcaptionskip{-1ex}
 \includegraphics[scale=0.025,]{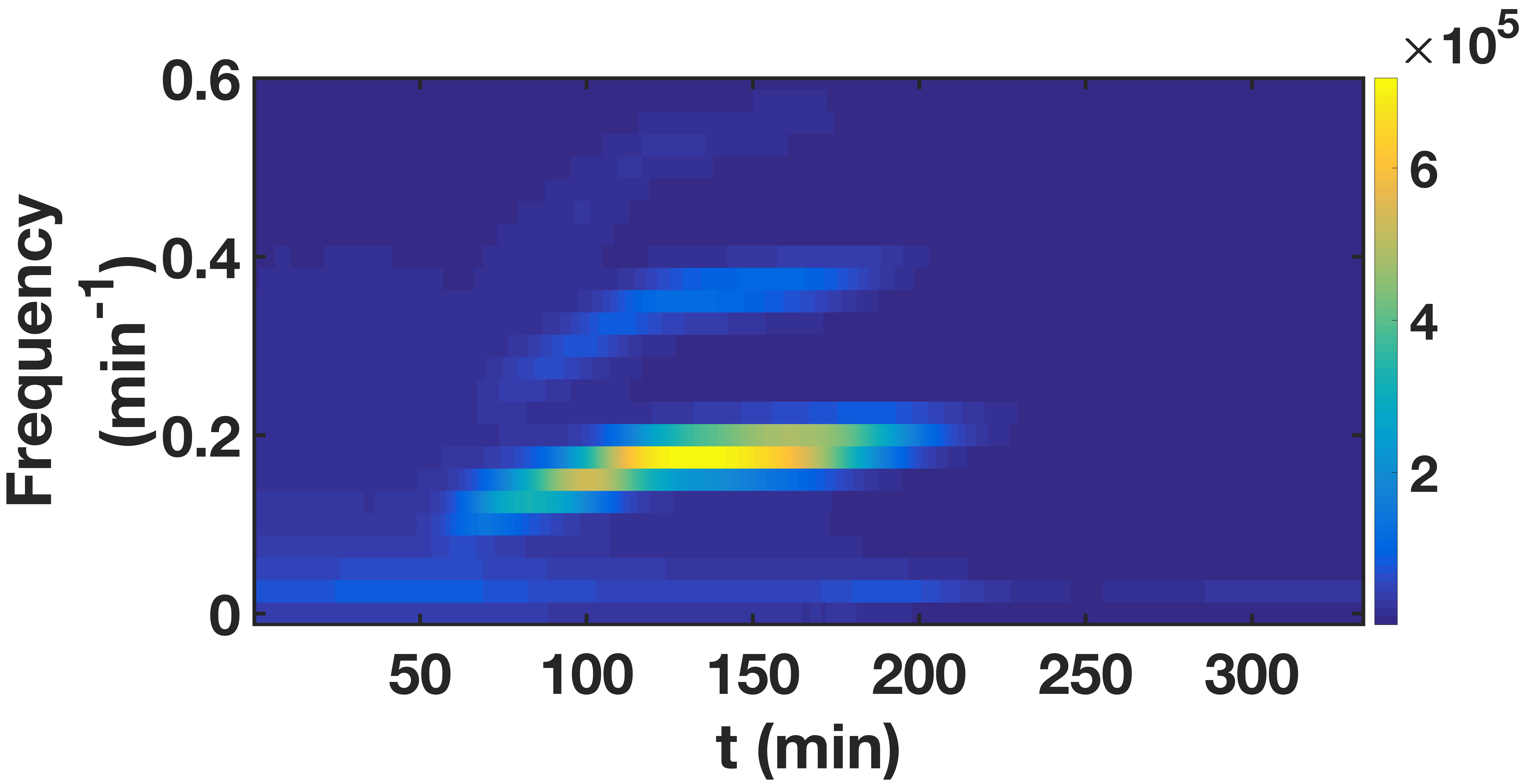}%
 \caption{Sample spectrogram obtained from an experiment. The slight increase in frequency (the y-axis) with observation time t (the x-axis), mentioned in the results section, is visible. The colormap shows the amplitude of the power spectrum. In this case, it has a maximum at around 140 $min$. The power spectrum at this time is taken, its peak fit with a quadratic polynomial and the frequency at the maximum amplitude found. \label{spectrogram}}
 \end{figure}
\subsection{Data analysis}
\subsubsection{Oscillation period}
First, the dark field images were filtered in space, only allowing structures between $3 \ mm$ and $0.2 \ mm$ to pass through. From these processed image stacks, we obtained the time series of the intensity for each pixel with a window of 40 $min$ (corresponding to about 6 oscillations). Then, we took the Fourier transform of these time series. The power spectra of the Fourier transforms thus obtained were spatially averaged giving one power spectrum for each time point. For example, we started with the first image of the image stack and found the time series of the intensity for all the pixels by collecting their intensity values  from images corresponding to the next 40 $min$. Then, we took the Fourier transform of each of these time series and found their power spectra. These power spectra were then averaged to get one average power spectrum corresponding to the first image. Then we moved to the second image in the image stack. Again, we found the intensity time series of each pixel by collecting its intensity values from images corresponding to the next 40 $min$. As before, we took the Fourier transform of each of these pixels and found their power spectra. The spatially averaged power spectrum corresponds to the power spectrum of the second image. This procedure was repeated resulting in a power spectrum for each time point. The magnitude of these power spectra were plotted over time to give a spectrogram. An example spectrogram from an experiment is shown in figure \ref{spectrogram}. From the spectrogram, we found the time at which the power spectrum has the highest amplitude. This time corresponds to the time when the patterns are well developed. We fit the peak of the power spectrum at this point with a second degree polynomial to obtain the frequency corresponding to the maximum and converted this frequency to period of oscillations. In some cases, e.g. at large initial starvation times, only 2 or 3 oscillations of the targets were observed. In these cases, because it was not possible to use the Fourier transform technique, we obtained the period by finding the distance between peaks in the temporal intensity profile. The patterns formed in the simulations were analyzed with the same procedure as above. Here however, the limits of the band pass filter were $1.5 \ mm$ and $0.1 \ mm$ because the wavelengths are smaller.
\begin{figure}
\setlength\belowcaptionskip{-1ex}
 \includegraphics[scale=0.35,]{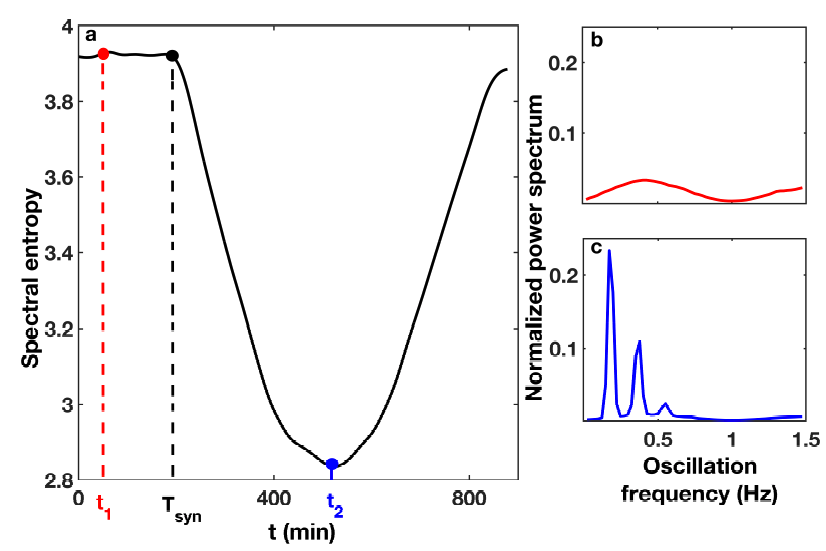}%
 \caption{(a) The temporal variation of spectral entropy obtained from an experiment. The point of maximum curvature, defined to be the onset of synchronization, is denoted by $T_{syn}$ (b) The normalized power spectrum corresponding to time $t_1$. It has a non-zero power for almost all frequencies. Using equation (\ref{entropyeq}), the spectral entropy at this point is 3.92. (c) The normalized power spectrum corresponding to time $t_2$. The power is high only for a few frequencies and zero for all others. The spectral entropy is calculated to be 2.85.\label{specentropy} }
 \end{figure}
\subsubsection{Onset of synchronization}
We define the onset of synchronization using the spectral entropy (as described in \cite{specentropy}). We found the spatially averaged power spectrum at each time point of the data, as described before. To find the onset of synchronization, we first normalized these power spectra and used the following definition of spectral entropy, S:
\begin{equation}\label{entropyeq}
S=-\sum p_{i}log(p_{i}).
\end{equation}
Here $p_i$ is the amplitude of the $i^{th}$ frequency of the normalized power spectrum. If a time series represents white noise, all frequencies are equally dominant ($p_i$ is high for all the $i$ frequencies) and this spectral entropy is high (figure \ref{specentropy}b). This is the case in the initial images, where no patterns occur. If the signal comes from a single frequency source, the value of the spectral entropy is lower because $p_i$ is close to zero for almost all frequencies except one. (figure \ref{specentropy}c) This occurs when all the pixels or cells oscillate at the same frequency. So, the spectral entropy has a high value for unsynchronized noisy states and a low value for synchronized states. A plot of spectral entropy as a function of time shows the emergence of the synchronized state, when the spectral entropy begins to decrease (figure \ref{specentropy}a). We find the minimum of the second derivative of this spectral entropy, which gives the point of maximum curvature, and define it to be the onset of synchronization. The same procedure was performed for patterns in both experiments and simulations.
\subsubsection{Spiral density}
To find the spiral density, we began with the Fourier transform of the time series of the intensity of each pixel in the images, for a window of 40 $min$. After spatially averaging the power spectrum at each time point, we found the dominant frequency at each time point. We defined the phase at each pixel to be the phase of the oscillation at the dominant frequency (obtained from the real and imaginary parts of the Fourier transform). A map of this phase, ranging from from $-\pi$ to $\pi$ for all the pixels is the phase map of that image. A spiral core is a phase singularity, a point where phase is not defined. (For more on phase singularities, see \cite{winfreebook}.) One way to detect phase singularities is by integrating the gradient of the phase along a closed curve. Such an integral is non-zero only if the loop encloses a singularity. 
\begin{equation}\label{wn1}
n.2\pi = \oint \vec{\nabla \phi} .\vec{dl} = \oint \vec{k} .\vec{dl}
\end{equation}
Here $\vec{k}$ is the wave vector and $n$ is an integer. Stokes\ theorem gives
\begin{equation}\label{wn2}
n.2\pi = \int (\vec{\nabla} \times \vec{k}).\vec{dA}.
\end{equation}
\begin{figure}
\setlength\belowcaptionskip{-1ex}
 \includegraphics[scale=0.2]{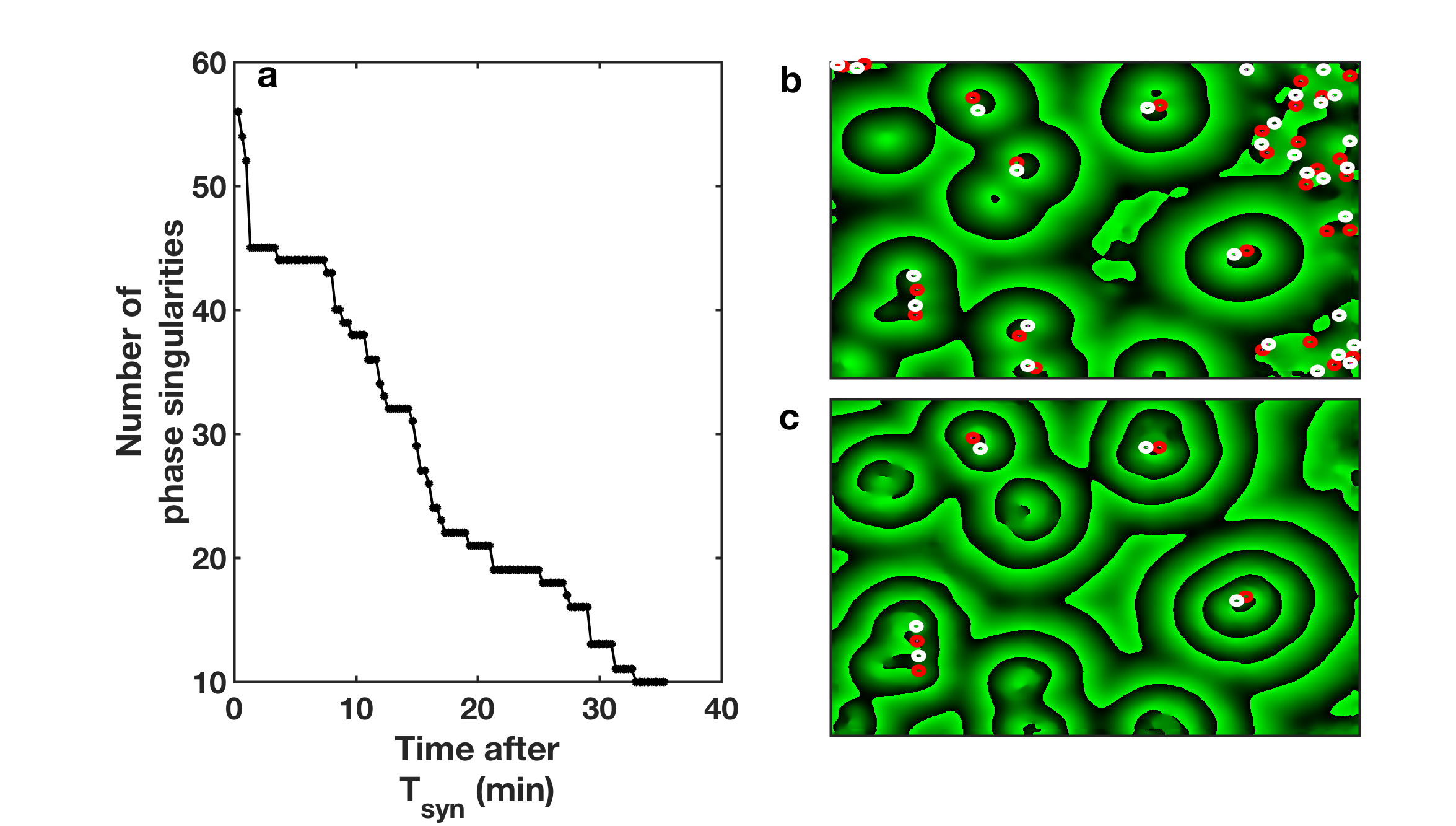}%
 \caption{The result of our counting algorithm. The left panel shows the decrease in the number of phase singularities counted. Initially a lot of phase singularities are detected due to noise (top right panel). However, with time the number of singularities that persist decreases. At the end of the counting, the spurious singularities are discarded (bottom right panel)\label{phsingvstime}}
 \end{figure}
The cross product in equation (\ref{wn2}) was evaluated using the kernel method described in \cite{Bray} to obtain the singularities in our patterns. For spirals, the winding number $n$, is either +1 or -1. Noise introduces spurious singularities. To eliminate such spurious singularities, we spatially filtered the phase map by first converting the phase to a complex number by raising it to an exponential of $e$ and multiplying with $i$. In this complex plane, we filtered in 2D using a box-average filter. The filtered phase map was then obtained from the angles of the resulting complex numbers. \cite{Janthesis}

However, this was not enough to remove the spurious singularities. So, we defined spirals to be the singularities that persisted for 40 $min$. To achieve this, we detected all the singularities in the first image, defined a tolerance length, and used the fact that the spiral cores are never very close to one another. In the next frame, we checked if the singularities of the first image persisted in a box around the original position, with side equal to the pre-defined tolerance length. If no singularity was detected in this box, the original singularity was discarded. Also, if two singularities were closer than a pre-defined distance, they were discarded. This process continued for all the frames involved in counting. The counting of the phase singularities started about 20 $min$ after $T_{syn}$, when the patterns are well established.  Figure \ref{phsingvstime} shows an example output of our algorithm. In figure \ref{phsingvstime}b, we see the phase map of the first image considered for counting the singularities. The red and the white rings denote the +1 and -1 winding numbers respectively. Despite the filtering, it has many spurious singularities. Figure \ref{phsingvstime}c shows the phase map of the last image considered for counting singularities. We see that all the spurious singularities have been eliminated by the tracking algorithm. Figure \ref{phsingvstime}a shows the decrease in the number of singularities with time. 

As described earlier, for the experiments and simulations with large initial starvation times, the signal to noise ratio is very poor, resulting in numerous spurious singularities even when no spirals are visible. The method just described fails for these cases. Therefore, to disregard such cases, we found the actual amplitude of the signal by finding the absolute value of the Fourier transform at each time point for the data. We compared the maximum values of the signal for all experiments. As expected, there was a clear decrease in the signal amplitude at $7 \ h$ of initial starvation time. We set this value as a cut-off. Any experiment (single population and population mixtures) that had a lower signal amplitude than this cut off was classified as forming targets.

\section{ACKNOWLEDGEMENTS}
The authors thank Dr. Noriko Oikawa and Dr. Albert J Bae for their programs for the data analysis. Further, the authors thank Ms. Katharina Gunkel, Ms. Maren S M{\"u}ller and Ms. Tina Althaus for their cheerful help with the preparation of the cells. The authors also thank Dr. Jan Christoph for suggesting the filtration method to detect phase singularities. This work has been supported by the Max Planck Society. A.G. acknowledges the MaxSynBio Consortium which is jointly funded by the Federal Ministry of Education and Research of Germany and the Max Planck Society.
\medskip
\bibliography{ref}

\end{document}